\documentstyle[aps, draft, epsf, floats]{revtex}

\def\gesim{\,{\raise-3pt\hbox{$\sim$}}\!\!\!\!\!{\raise2pt\hbox{$>$}}\,}
\def\lesim{\,{\raise-3pt\hbox{$\sim$}}\!\!\!\!\!{\raise2pt\hbox{$<$}}\,}
\def\pr{G}
\def\pd{g}
\def\ppr{{\bf\pr}}
\def\ppd{{\bf\pd}}
\def\pmb#1{\setbox0=\hbox{#1}\kern-.025em\copy0\kern-\wd0
  \kern.05em\copy0\kern-\wd0 \kern-.025em\raise.0433em\box0 }

\newcommand\beq{\begin{equation}}
\newcommand\eeq{\end{equation}}
\newcommand\bea{\begin{eqnarray}}
\newcommand\eea{\end{eqnarray}}
\def\so{\sum_{|\omega|>\Omega}}
\def\soop{\sum_{|\omega|,|\omega'|>\Omega}}
\def\Psit{\hat\Psi}
\def\Ht{\hat H}
\def\psit{\hat\psi}
\def\chit{\hat\chi}
\def\com#1#2{{\left[#1,#2\right]}}
\def\slow#1{\left\langle{#1}\right\rangle}

\def\bit{\eta}
\def\inv#1{{1\over#1}}
\def\ucal{{\cal U}}
\def\hef{{H_{\rm eff}}}
\def\half{\inv2}
\def\im{{\hbox{\bf Im}}}
\def\heft{\hat\hef}
\def\sigbf{{\pmb{$\sigma$}}}
\def\pcal{{\cal P}}
\def\ccal{{\cal C}}
\def\rr{{\bf r}}
\def\avg#1{{\left\langle#1\right\rangle}}
\def\up#1{^{(#1)}}
\def\kcal{{\cal K}}
\def\ecal{{\cal E}}
\def\ocal{{\cal O}}
\def\LL{{\bf L}}
\def\ff{{\bf f}}

\def\aa{{\bf a}}
\def\zz{{\bf z}}
\def\xx{{\bf x}}
\def\re{{\bf Re}}
\def\nubf{{\bf\nu}}

\begin{document}

\preprint{UCRHEP-T294}

\title{Effective theory of systems coupled strongly to rapidly-varying
external sources.}
\author{R. Huerta$^a$ and J. Wudka$^b$}
\address{$^a$ Departamento de F\'\i sica Aplicada, Cinvestav-IPN
Unidad M\'erida.\\ M\'erida, Yucat\'an 97310, M\'exico\\
$^b$ Department of Physics, University of California, Riverside CA 92521-0413.}
\date{\today }
\maketitle

\begin{abstract}
We consider quantum systems which interact strongly with a 
rapidly varying environment and derive a Schr\"odinger-like equation 
which describes the time evolution of the average wave function. 
We show that the corresponding Hamiltonian can be taken to be Hermitian
provided all states are rotated using an appropriate unitary
transformation. The formalism is applied to a variety of systems and
is compared and contrasted with related results describing stochastic  
resonances.
\end{abstract}

\draft
\pacs{PACS: 05.40, 14.60.P, 42.15, 32.80, 05.10.G}

\section{Introduction}
\label{sect:intro}

The study of quantum systems which interact strongly 
with their environment often presents
serious challenges due to the possibility that these interactions cannot
be neglected and, in addition, also vary rapidly and randomly with time~\cite{fast}.
The effects of such external fields is often unavoidable
and interesting and can lead to unexpected phenomena
such as, for example, those studied under the blanket term of stochastic 
resonances~\cite{sr}. In this paper we will study one subset of such systems.

We will assume that the interactions with the environment are described
by a time-dependent contribution to the Hamiltonian denoted by $H'(t)$ which
cannot be treated perturbatively. In analogy with a similar
problem in mechanics~\cite{kapitsa} we will assume self-consistently
that the states of the system can be decomposed into a sum of
slowly varying modes and  high-frequency components of small amplitude. 
Using this decomposition we will show that
the time evolution of the slow modes is determined by an
effective Hamiltonian which, to leading order, depends quadratically on
the external interaction $H'$.
The formalism assumes that the time scales associated with the
interactions with the environment are much shorter than all other
frequencies in the problem.  Denoting by $ \Omega $ a typical
frequency of the interaction $H'$, we will obtain a solution to
Schr\"odinger's equation as a series in $ 1/\Omega $. 

The effective Hamiltonian describes the average time evolution of the
system and can exhibit resonances under some special circumstances which
will be illustrated using simple examples. 
It is also worth noting that the same formalism can be applied to {\em
any} system evolving according to a Schr\"odinger-like equation assuming
that the operators corresponding to the Hamiltonian contain terms which
vary rapidly in the evolution parameter. We also provide
examples of this type of generalization: using geometrical optics
we study light-ray propagation in a
random media, and, we determine the effects of a time-independent potential
which varies rapidly with position on the wave functions of a non-relativistic
particle.

The paper is organized as follows: in section \ref{sect:formalism}
we give a description of the formalism and find the effective
Hamiltonian that will be used in the applications. The behavior of the effective
Hamiltonian under unitary transformations is studied in 
section \ref{sect:unit.transf}; the formalism is then applied to 
various illustrative examples in section \ref{sect:exa}.
In section \ref{sect:fp} we give an alternative view of the problem in 
terms of the Fokker-Planck equation and the results are then
compared and contrasted with the formalism used in deriving the
standard stochastic resonances (section \ref{sect:compare}). Paring comments and conclusions
are presented in section \ref{sect:concl}. Finally, a mathematical detail is
relegated to the appendix. 
 
\section{Quantum systems with rapidly-varying external fields}
\label{sect:formalism}

We consider a generic quantum system with a Hamiltonian
of the form
\beq
H = H_0 + H',
\label{eq:htot}
\eeq
where $H'$ is time dependent with characteristic frequencies assumed
larger than all the other energy scales in the system (we take units
where $ \hbar=1$). In general we will also allow $H_0$ to vary with
time, but with the restriction that the time scale(s) associated with $H_0$
are much smaller than those associated with the time variation of
$H'$. In addition we assume that $H'$ is larger than $H_0 $ so that,
symbolically
\beq
H_0, ~ \dot H_0/H_0 < H' < \dot H'/H'.
\label{eq:cond}
\eeq

More specifically we assume that $H'$ admits a Fourier expansion of the type
\beq
H' = \so H_\omega e^{-i\omega t},
\label{eq:hpfe}
\eeq
where the sum is over a set of frequencies $ \{\omega\}$ such that the
differences also obey  $ | \omega - \omega' | \ge \Omega $. In general
we will allow the Fourier coefficients $ H_\omega $ to be time-dependent,
but, as for $H_0 $, we assume that the corresponding frequencies are
small compared to $ \Omega $. Henceforth ``slow'' will mean
``of frequency $ \ll \Omega $''.

In solving the Schr\"odinger equation for such a system we will assume
that the wave function can be separated into a slowly varying piece
$ \psi $ and a rapidly varying (frequency $\gesim \Omega$)
``ripple'' $ \chi $ of small amplitude,
\beq
 \Psi = \underbrace{\psi}_{\rm slow} + \underbrace{\chi}_{\rm fast}.
\label{eq:psitot} 
\eeq
We will then match slow and fast terms, noting that even though
$ | \chi | $ is small, $ \dot \chi$ can be large.

We will assume that all quantities can be written as the sum of 
a slowly varying piece (containing frequencies $ \ll \Omega $) and a
fast piece (in special cases one or the other may vanish). 
It then proves convenient to  introduce the following notation: for any
quantity $ \Xi$
\bea
\slow\Xi:&&\hbox{slow~part~of~}\Xi .
\label{eq:sf}
\eea
For example, $ \slow\Psi = \psi$, $\slow H = H_0$, $\slow{H'}=0 $.

To solve Schr\"odinger's equation for this class of systems we
consider a typical term in (\ref{eq:hpfe}) and define the expansion
parameter
\beq
\bit \sim { |H_\omega| \over \Omega }
\label{eq:defofbit}
\eeq
(alternatively $ \bit \sim | \int dt H' | $); the previous restrictions
imply that $ \eta < 1 $. We will now
assume that the wave function has an expansion in powers of $ \bit $:
\beq
\Psi = \psi + \chi_1+ \chi_2 + \cdots , \quad \chi_n \sim \bit^n.
\label{eq:expofpsi}
\eeq
This expansion is useful for $ n \ll 1/\eta $, beyond this order we
typically obtain $n-$fold products of slowly varying quantities which
can generate terms with frequency $ \sim\Omega$, and the separation
between slowly and rapidly varying terms cannot be maintained. These
effects are, however, subdominant since their {\em
amplitude} are suppressed by a small factor $ \propto \eta^n \ll1 $.

Substituting $H$ from (\ref{eq:htot}) and $ \Psi $ from (\ref{eq:psitot}) in 
Schr\"odinger's equation
\beq
H\Psi = i \dot \Psi,
\eeq
we find, to lowest order in $ 1/\Omega $
\beq
i \dot \psi + i\dot\chi_1 + O(\Omega \bit^2 )
= 
H_0 \psi + H' \psi  + O(H_0 \bit, \Omega \bit^2 ),
\eeq
whence
\beq
i \dot \chi_1 = H' \psi , \qquad i\dot \psi  = H_0 \psi .
\eeq
Since $\psi$ is slowly varying, the first equation can be solved
 to this order in $ \bit $ by taking $ \psi $ constant, namely,
\beq
\chi_1 =  \ucal \psi; \qquad
\ucal = \left(\inv i\int dt H'\right)  = \so \inv\omega H_\omega e^{-i \omega t }.
\label{eq:chione}
\eeq

To the next order we write
\beq
\Psi = \psi + \ucal \psi + \chi_2 + O(\bit^3),
\eeq
and obtain
\beq
i \dot\psi + i \ucal \dot\psi + i \dot\chi_2 
=
H_0\psi + H_0 \ucal\psi + H' \ucal\psi+O(\Omega \bit^3, H_0 \bit^2 ).
\eeq
Note that $H'\ucal$ contains both slow and fast terms. Using the notation
(\ref{eq:sf}) we find
\bea
i \dot \psi &=& \left( H_0 + \slow{H'\ucal} \right) \psi, \cr
i \dot \chi_2 &=& \left( [H_0,\ucal] + H'\ucal - \slow{H'\ucal} \right) \psi,
\label{eq:eqforpsi}
\eea
where the  second equation can be solved (to the present order in $ \bit $) by
neglecting the time variation in  $H_0$ and $ \psi $. 

To this order the {\em average} wave function then obeys a
Schr\"odinger-like equation with an effective Hamiltonian~\cite{v.w}
\beq
\hef = H_0 + \slow{H'\ucal}.
\label{eq:hi}
\eeq

It is easy to see that to this order in $ \bit$ $ \hef$ is Hermitian,
however, to order $ \bit^2 $ we find
\beq
\hef =  H_0 + \slow{H'\ucal} - \slow{\ucal
 \left( \com{H_0}{\ucal} + H'\ucal \right)},
\label{eq:hscnd}
\eeq
which is not Hermitian:
\bea
\hef-\hef^\dagger &=& 
\slow{ H' \ucal + \ucal H'  + \com{\ucal^2}{H_0} }+ \cdots. \cr
&=& i \partial_t \slow{\ucal^2} + \com{\slow{\ucal^2}}{H_0} + \cdots
\label{eq:hscnd.nh}
\eea
and equals, to this order the {\em total} time derivative of the
operator $ \slow{\ucal^2} $.
This property corresponds to the fact that there is some probability 
``leakage'' of order $\bit^2 $ from the slowly varying
part of the wave function to the rapidly varying ripple. This is to be
expected since $\slow{|\chi|^2} = O(\bit^2) $ and is non-zero in
general.

The non-Hermiticity of the effective Hamiltonian for the slowly varying
modes frequently appears in expansions similar to the one considered here
\cite{bur}~\footnote{In \cite{bur} a non-Hermitian term was found
already in the first order, the discrepancy between this result and the one
obtained here is due to different assumptions concerning the
time-dependence and magnitude of the various terms in the Hamiltonian,
leading to different expansion parameters.}.
This result can be better understood by considering the
behavior of the above expansion under unitary transformations
to which we now turn.

\section{Unitary transformations}
\label{sect:unit.transf}

In this section we determine the behavior of the effective Hamiltonian
(\ref{eq:hscnd}) under unitary transformations. We show below that  the
non-Hermitian piece in (\ref{eq:hscnd}) is modified under such
transformations and, in fact, can be completely eliminated.

For the case of a
constant transformation, $ \Psi \to \ccal \Psi $ with $ \dot \ccal =0 $
it is clear that $ H \to \ccal^\dagger H \ccal $ and 
$ \hef \to \ccal^\dagger \hef \ccal $. 
If the unitary transformation is time-dependent, however, the result is
more complicated. We will concentrate on transformations of the form
\beq
\Psi = e^F \Psit,
\eeq
where $F$ is anti-Hermitian, rapidly varying, 
and of order $ \bit $. The Hamiltonian for the transformed states 
$\Psit $ is 
\bea
\Ht &=& e^{-F} H e^F -i e^{-F} \partial_t e^F \cr 
&=& H +\com H F + \half \com{\com H F}F - i \dot F 
   - {i\over2}\com{\dot F}F-{i\over6}\com{\com{\dot F}F}F + O(\bit^3).
\eea

A tedious repetition of the procedure outlined in
section \ref{sect:formalism} gives the following expression for the 
corresponding effective Hamiltonian
\bea
\Ht_{\rm eff} &=& H_0  + \slow{H' \ucal} + {i\over2} \partial_t
\slow{F(F-2\ucal)}+\half \slow{\com{H_0}{\com F \ucal}} \cr 
&& +  \half \slow{\com{\com{H_0}\ucal}\ucal}- \slow{\ucal H' \ucal}
-\half\slow{\com{H_0}{(F-\ucal)^2}}+O(H_0 \bit^3, \Omega \bit^4).
\eea

For a general choice of $F$ this expression is still non-Hermitian. However
for the special case
\beq
F =  \ucal+ O(\bit^2),
\eeq
we obtain
\bea
\Ht_{\rm eff} &=& H_0  + \half \slow{\com{H'}\ucal}+
\half \slow{\com{\com{H_0}\ucal}\ucal}- \slow{\ucal H' \ucal}+
O(H_0 \bit^3, \Omega \bit^4),
\label{eq:theh}
\eea
which is explicitly Hermitian and, in fact, it is identical to the
Hermitian part of (\ref{eq:hscnd}).
It is, of course, always possible to return to the original frame using $
\Psit = \exp(-\ucal + \cdots ) \Psi $.

The wave function in the new frame, $ \Psit$, has an expansion similar to
(\ref{eq:expofpsi}) 
\beq
\Psit = \psit + \chit_1+ \chit_2 + \cdots , \quad \chit_n \sim \bit^n,
\label{eq:expofpsit}
\eeq
where the slowly varying piece $\psit$ evolves
unitarily in time since (\ref{eq:theh}) is Hermitian (at least
to order $ \bit^2 $). Using (\ref{eq:chione}) we find
\bea
\psit &=& \left(1-\half \slow{\ucal^2} \right)\psi + O(\bit^3) ,\cr
\chit_1 &=& 0.
\eea
The $O(\bit^2) $ difference between $ \psi $ and $\psit$
quantifies the probability leak into the rapidly
varying sector for the original frame. The second order term in the wave
function $ \chit_2 $ cannot be determined without a specific choice for
the $ O ( \bit^2 )$ terms in $F$. The specific form of the relation
between $ \psit $ and $ \psi $ can also be understood form the
expression for the non-Hermitian part of the effective Hamiltonian
obtained in (\ref{eq:hscnd.nh}).

We conjecture that this procedure can be carried order-by-order in $ \eta
$, but since we will not need these higher order corrections
we will not pursue this further.
The expression (\ref{eq:theh}) is the form of the effective Hamiltonian
which will be used in the following examples.

\section{Examples}
\label{sect:exa}

The previous results can be applied to a variety of systems.
In this section we will consider 5 such examples. Our main concern will
be to illustrate a wide range of systems that can be studied using the
above formalism

\subsection{$N$-level quantum systems}
\label{sect:exa.n.level}
In this section we consider a quantum system with a finite number of
states. This serves, for example, as a model for spin or flavor changes
in elementary particles; it also describes the basic physics of nuclear
magnetic resonance and related phenomena.
The most general Hamiltonian for these systems can be expanded in
terms of the generators of $SU(N)$ which we denote by $ \{ \lambda^a\}$
and which satisfy
\beq
\com{\lambda^a}{\lambda^b} = i C_{abc} \lambda^c.
\eeq
We can then write 
\beq
H_0 = \sum_a f^a \lambda^a, \qquad H_\omega = \sum_a \pd_\omega^a ,
\lambda^a
\eeq
where $H_\omega$ are the Fourier coefficients of the rapidly-varying
Hamiltonian (see (\ref{eq:hpfe})). Note that
$ \pd_{-\omega}^a{}^* = \pd_\omega^a $ since $H'$ is Hermitian.

Substituting in (\ref{eq:theh}) we find
\bea
\hef &=& \sum_a \varphi^a \lambda_a ,\cr
\varphi^a &=& f^a 
-i  \so { \pd_\omega^c{}^* \pd_\omega^b \over 2\omega}  C_{abc} 
+ \so { f^d \pd_\omega^c{}^* \pd_\omega^b \over 2  \omega^2} C_{dbe} C_{eca} 
+ \soop { \pd_\omega^b \pd_{\omega'}^c \pd_{\omega+\omega'}^d{}^* \over 3
		\omega'(\omega+\omega')} C_{dbe} C_{eca} + \cdots,
\label{eq:hN}
\eea
where the ellipsis denote higher order terms in $\bit$.

For the particularly simple case of a two-level system with $ C_{abc} =
2 \epsilon_{abc} $ and $ \lambda_a = \sigma_a $ (the usual Pauli
matrices) this reduces to
\beq
\hef = \left( \ff - \so {\ppd_\omega \times \ppd_\omega^* \over \omega }
 + \cdots \right) \cdot \sigbf .
\label{eq:twolevelcase}
\eeq

\begin{figure}
\centerline{\vbox to 4in{\epsfxsize=8 in\epsfbox[00 -250 612 542]{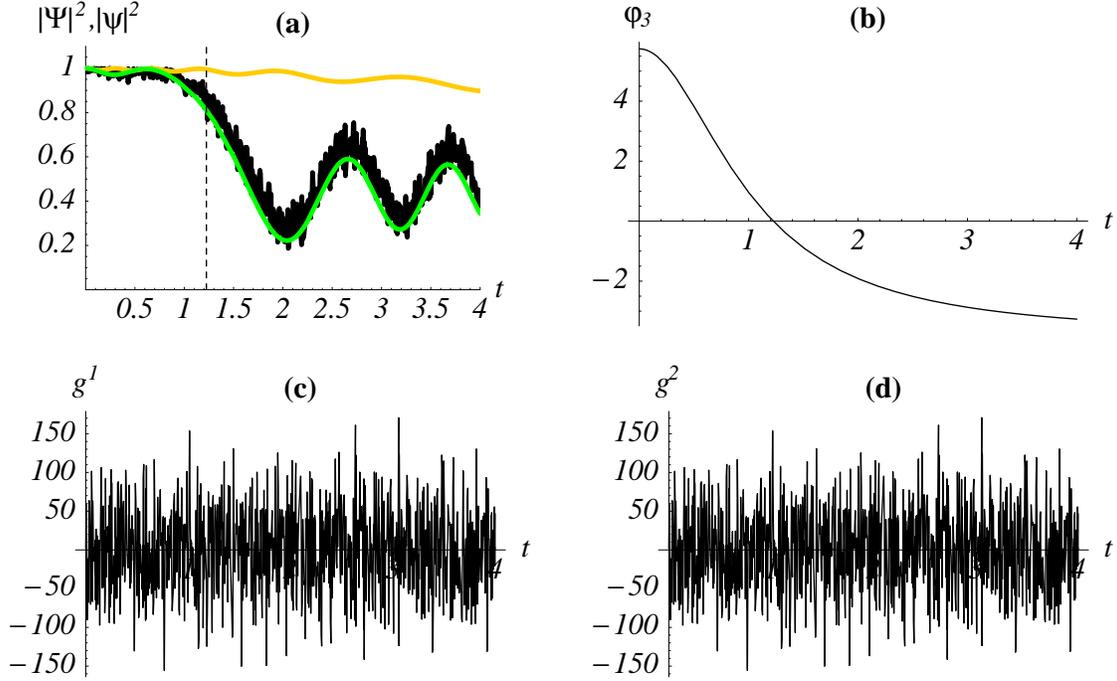}}}
\caption{ Example of a resonant phenomenon
induced by the presence of rapidly varying interactions of
large amplitude. {\bf(a):} comparison of $
| \Psi|^2 $ obtained by an exact (numerical)
integration of Schr\"odinger's equation (black jagged curve) with 
$ | \psi |^2 $ obtained integrating (\ref{eq:eqforpsi}) using the
effective Hamiltonian in (\ref{eq:twolevelcase}, \ref{eq:twolevl}) (light
superimposed curve). The top light curve is the result of integrating
Schr\"odinger's equation when $ H'=0 $; the dotted vertical line denotes the time
at which the diagonal elements in $ \hef $ vanish. 
{\bf(b):} Diagonal element of $ \hef $; for this
example we chose $ E'=4.78684 $ and $ f^3 = E' [2/(1+t^2)+1/5] $. 
{\bf(c,d):} Non vanishing elements of $H'$; the
specific expression used was
$ \pd^1 + i \pd^2 =$ $  
39.9567 e^{926.291 t-0.956304i} + $ $ 
35.6145 e^{984.461 t+0.660091i} + $ $ 
39.1024 e^{1057.84 t-0.732253i} + $ $ 
30.4239 e^{1208.99 t+2.43462 i} + $ $ 
29.1863 e^{1953.06 t+0.719083i} $. For this example, $ \eta\simeq0.03 $
}
\label{fig:fi}
\end{figure}

It is clear that there is the possibility for the $H'$-induced
terms to generate resonances if we allow the $f^a$ to vary slowly in time.
To see this explicitly we consider (\ref{eq:twolevelcase}) taking for
simplicity $ f^2=\pd^3_\omega =0 $, and $ f_1 = $constant. Substituting we find
\bea
\varphi_1 = f^1 , \qquad 
\varphi_2 = 0 , \qquad 
\varphi_3 = f^3 - E'+ O(1/\Omega^2 ),
\label{eq:twolevl}
\eea
where
\beq 
E' = 2 \so \inv\omega \im \pd^2_\omega{}^* \pd^1_\omega.
\eeq
If $f^3 $ is allowed to vary slowly with time
the effective Hamiltonian
will exhibit a resonance when $ f^3 = E'$
if  $ |E'| \gg |f^2| $. This resonance 
will lead to large transition amplitude provided 
$ f^3 $ varies sufficiently
slowly: $ |f^2| \gg \sqrt{|\dot f^3|} $ (which is the usual adiabatic
resonance condition \cite{messiah}). A specific example is 
presented in fig. \ref{fig:fi}.

In fig. \ref{fig:fi} we also compare the results obtained using the effective
Hamiltonian (\ref{eq:twolevelcase}) to those obtained solving the
Schr\"odinger equation exactly with initial conditions $ \Psi ( t = 0 )
= \psi ( t =0 ) = 1$.
It can be seen that the solutions
obtained using $\hef$ do indeed describe the average behavior of the
wave-function provided $ \bit$ is sufficiently small. We conclude that,
at least for the case of a 2-level system, the effective Hamiltonian
(\ref{eq:theh}) accurately describes the average evolution of the
system.

The presence of noise ({\it i.e.}, $H'$) can then
generate unexpected resonances. The condition for these to occur is,
qualitatively,
\beq
\Omega H_0 \sim H'{}^2;
\label{eq:res.cond}
\eeq
we will see later that these resonances are related, but not identical,
to the well-studied stochastic resonances \cite{sr}.

\subsection{$\delta$-function comb}
\label{sect:exa.comb}

A simple model where the above formalism can be applied and
which is also exactly solvable is provided by a 2-level system with
Hamiltonian
\begin{equation}
H = \sum_{n} \sum_{i=1}^N \Lambda_i \delta(t-t_i-nT); \quad 0< t_i<T,
\end{equation}
where the matrices $ \Lambda_i $ are Hermitian. This represents a set
of $N$ $\delta$ functions which repeats with period $T$.
This type of potential is of interest in signal processing~\cite{comb.signal} 
and is also similar to the one used in the study the effect of a laser 
beam on a set of charged particles~\cite{comb.laser}. 

For simplicity we will assume 
\begin{equation}
\Lambda_i = \pmatrix{0 & \lambda_i \cr \lambda_i^* & 0 } ; \qquad
\sum_{i=1}^N \lambda_i =0,
\label{eq:simplif}
\end{equation}
in this case $H_0=0, $ so that $ H = H'$, we will also take $ \Omega =
2\pi/T $.

In order to obtain the effective Hamiltonian we first construct
\begin{equation}
-i \int_0^t dt \; H' =-i  \sum_{i=1}^N \Lambda_i \Theta(t-t_i) , \quad (0 < t < T ), 
\end{equation}
where $\Theta$ denotes the step function. 
$ \ucal $ is then the fast part of this quantity,
\begin{equation}
\ucal = -i \int_0^t H' + i \slow{ \int_0^t H'} = -i\sum_{i=1}^N
\Lambda_i \left[ \Theta( t - t_i ) -1 + {t_i \over T } \right] , \quad (0 < t < T ), 
\end{equation}
where the slow part of a quantity is obtained by averaging over the
period $T$.

Using then (\ref{eq:simplif}) and substituting in (\ref{eq:theh}) we easily find 
\begin{eqnarray}
H_{\rm eff} &=& \half \slow{\com{H'}\ucal} + \cdots \cr
&=& \inv T \left[ \sum_{i>j} {\rm\bf Im}\lambda_i \lambda_j^* \right]
\sigma_3 + \cdots.
\label{eq:comb.heff}
\end{eqnarray}
For this system we have $ \Omega \sim 1/T $ so that $ \eta =
\hbox{max} \{ | \lambda_i | \} $; it then follows that (\ref{eq:comb.heff}) will be
accurate provided $ | \lambda_i | \ll 1 $.

This model can be solved exactly by elementary means. Replacing $
\delta (t - t_i - nT ) $ by a rectangle of height $ 1/\tau $ and width $
\tau $ centered at $ t = t_i + n T $ it is easy to see that in the
limit $ \tau \to 0 $
\begin{equation}
\Psi(t_i^+) = e^{-i \Lambda_i} \Psi(t_i^-),
\end{equation}
where $ t_i^\pm = t_i \pm \delta,~\delta\to0$. It follows that
\begin{equation}
\Psi(T) = e^{-i \Lambda_N}e^{-i \Lambda_{N-1}} \cdots e^{-i
\Lambda_1}\Psi(0).
\end{equation}
In the limit where the $ \lambda_i $ are small we obtain
\begin{eqnarray}
e^{-i \Lambda_N}e^{-i \Lambda_{N-1}} \cdots e^{-i
\Lambda_1} 
&=& \exp\left[ - \half \sum_{i>j} \com{\Lambda_i}{\Lambda_j} + \cdots \right] \cr
&=& \exp\left[ - i T H_{\rm eff} + \cdots \right] 
\end{eqnarray}
which shows that, at least for small $ \lambda_i, $ $ \hef $
determines the leading contributions to the average time evolution of the 
wave function.

\subsection{Geometrical Optics example}
\label{sect:exa.optics}

The calculations in the previous sections referred to quantum systems,
but it clear that any system whose dynamical equations can be cast into
a Schr\"odinger-like form can be treated in the same way. In particular
for this general case there is no need to require $H$ to be a Hermitian
operator. 

An example of this situation is provided by the description
of light-ray evolution within geometrical optics~\cite{h.and.z}. For small angles
the position and direction of light ray within geometrical optics can be 
described using a  two-component vector 
\beq
\pmatrix{ h \cr \alpha },
\eeq
where $h$ denotes the height with respect to a reference line
and $ \alpha $ the tilt (assumed to be small). Any transformation of a
light ray can be described using a $ 2\times2$ matrix~\cite{h.and.z}. In particular
\bea
\hbox{translation}: && \pmatrix{1 & x \cr 0 &1},\cr
\hbox{refraction}: && \pmatrix{1 & 0 \cr (n_1/n_2-1)/R & n_1/n_2},
\eea
where the translation is by a distance $x $ and the refraction
is from a medium of refraction index $ n_1 $ to another with index
$ n_2 $ and $R$ denotes the radius curvature of the interface.

We now assume that $R$ and the index of refraction change
smoothly though rapidly with distance. We define
\beq \nu = - \inv n { d n \over dx},
\eeq
 so that the general matrix which transports a ray by a distance $ \delta x $
is
\beq
M = \pmatrix{ 0 & 1 \cr \zeta' & \nu}.
\eeq
where primes denote $x$ derivatives and
\beq
\zeta = \int dx {\nu\over R}
\eeq
The system then corresponds to a two-level quantum system with
``time'' $x$ and ``Hamiltonian'' $ H = i M $. A simple application of 
(\ref{eq:hi}) yields
\beq
\hef = i \pmatrix{0&1\cr K^2 & 0 }, \qquad
K^2 = \slow{\zeta' ( \ln n) } = +\half  \slow{(\ln n)^2 (1/R)'},
\label{eq:optics.hef}
\eeq
where we assumed $ \slow{ \zeta'} = \slow{\nu} = 0 $ and we
kept only the first corrections induced by the
rapidly varying terms~\footnote{The second
expression for $K^2$ follows from
$  2 \nu (\ln n)/R = (1/R)'( \ln n)^2- [(\ln n)^2 /R]' $ and
$\slow{[(\ln n)^2 /R]'} = 0 $ to first order in the rapidly varying
quantities.}. Note that $K^2$ can be negative and that it vanishes
(at least to lowest order) when $R$ or $n$ are constant.

The effective operator which determines the translation over a finite
distance $X$ is then (assuming for simplicity that $K$ is position independent)
\beq
A = e^{-i X \hef} = \pmatrix{\cosh(K X)& (1/K)\sinh(K X) \cr
K \sinh(K X) & \cosh(K X) }
\eeq
(the $ K^2< 0 $ case is obtained by analytic continuation).
This matrix is equivalent to a thick symmetric lens with  
radius of curvature $\bar R$ and thickness $ \bar d$ such that
\beq
\inv{\bar R} = { K \over 1-\bar n} \tanh(K X/2) \qquad
\bar d = { \bar n \over K} \sinh( K X),
\eeq
where $ \bar n$ denotes the index of the lens material.

Thus, within the approximations inherent to geometrical optics,
the effects of a region of rapidly varying index
of refraction and curvature on light rays are equivalent, {\em on average} 
to those of a thick
lens of appropriately chosen characteristics.
For example, using a thick lens with a high index of refraction, $ \bar
n \gg 1 $ we find
\beq
K |K|= -{ 2 \bar n^2 \over \bar R  \bar d }.
\eeq
Conversely, a thick lens can be found which completely cancels the
effects of $A$ and this can be used to
measure the fluctuations in the original system
(more specifically, those fluctuations which contribute to $\zeta$).

The above expressions suffer corrections form higher-order terms in the
expansion in powers of $ \bit $. Using (\ref{eq:hscnd})~\footnote{As
mentioned previously there
is no reason to demand Hermiticity in the effective Hamiltonian for this
case} we find that the next-order term in (\ref{eq:optics.hef}) is
\beq
i \pmatrix{ 0 & 0 \cr
k^2 & 0 }, \quad k^2 = 
\slow{\zeta^2 } - \slow{\zeta}^2 
-{1\over2}\slow{\ln n  \left[ \ln n - {1\over2} \slow{\ln n} \right]\zeta'},
\eeq
where we assumed for simplicity that $ \slow{ (\ln n - \slow{\ln n})\zeta}$
and $ \slow{ [ \ln n - \slow{\ln n}]^2} $ are independent of $x$
(in general the averaged quantities may still vary slowly with $x$). The
second order correction is negligible provided $ | K^2  |\gg | k^2 | $. 

In addition there are
deviations form these predications due to the inherent limitations of
geometrical optics (for example, it is assumed the light
rays lie on a plane, diffraction is neglected, etc.). In neglecting them we
have tacitly assumed that the scale of all fluctuations is large
compared to the wavelength and that all reflection and refraction angles
are small.

\subsection{The noisy Jaynes-Cummings model}
\label{sect:exa.jc}

In this example we consider a simplified version of an atom 
interacting with a photon field, as  described by the Jaynes-Cummings
model~\cite{jc}, with the addition of two types of interaction 
with an external rapidly varying fields. We will show that this
problem is also well suited for study using
the techniques introduced above. We first assume that the external field are
coupled to the photons, and then directly to the atoms. We then
show that both situations are unitarily equivalent.

The unperturbed Hamiltonian for this model is
\beq
H_0 = \omega_0 a^\dagger a + \half \Omega_0 ( a^\dagger \sigma_-
+ a \sigma_+ ) + \epsilon \sigma_3 .
\label{eq:j.c.ho}
\eeq
Denoting by $ | n ; \uparrow \rangle $ and $| n; \downarrow \rangle $ the
states with $n$ photons and atomic spin up and down respectively, the
 eigenstates of $H_0$ are~\cite{jc} 
\bea
| n \pm\rangle &=&  c_{n\mp} | n;   \downarrow \rangle \pm 
\hbox{sign}(\kappa) c_{n\pm} | n-1; \uparrow   \rangle ,\cr
c_{n\pm} &=& \inv{\sqrt{2}} \left(1 \pm \inv{\sqrt{1+ n
\kappa^2}}\right); \qquad
\kappa = { \Omega_0 \over 2 \epsilon - \omega_0 },
\eea
with energies
\beq
E_{n\pm} = \left( n - \half \right) \omega_0 
\pm \left| \epsilon - \half \omega_0 \right| \sqrt{ 1 + n \kappa^2 } .
\eeq

\paragraph{Noisy photon field}

We will now couple the photons to external sources which
vary rapidly with time. The interaction Hamiltonian is assumed to be
\beq
H' = \xi a^\dagger + \xi^* a .
\label{eq:firstnoise}
\eeq
Substituting (\ref{eq:j.c.ho}) and (\ref{eq:firstnoise}) in (\ref{eq:theh}) yields
\beq
\heft = H_0 + \omega_0 \slow{|\theta|^2} - \im \slow{\theta^* 
\dot \theta} + O ( \bit^3) ; \qquad \xi = i \dot \theta.
\eeq
The difference between $\hat\hef$ and $H_0$ is, in this case,
trivial and can be eliminated by a simple change
in the overall phase of the states. 

Non-trivial
terms may arise at higher orders but this would require calculating the
(Hermitian version of the) effective Hamiltonian to order $\bit^3 $.
Instead of following this uninspiring approach we consider a
different way of introducing the interaction with the external fields
and then show that the corresponding effective operator corresponds to
the order $ \bit^3 $ contribution generated by (\ref{eq:firstnoise}). 
To this end we first consider a
unitary transformation of the form
\bea
S &=& \exp\left[ a^\dagger \zeta(t) - a \zeta^*(t) \right] ; \qquad
 i \dot \zeta - \omega_0 \zeta + \xi =0. 
\eea
Then the transformed Hamiltonian is
\bea
H_{\rm new} &=& S (H_0 + H' ) S^\dagger + i \dot S S^\dagger \cr
&=&H_0 - \half \Omega_0 \left( \sigma_- \zeta^* + \sigma_+ \zeta \right)
-\half \im( \zeta^*\dot\zeta) - \omega_0 | \zeta|^2 \cr
&=& H_0 + H'_{\rm new},
\label{eq:uth}
\eea
which defines $H'_{\rm new}$.
This shows that the original system is equivalent to one where the
external sources are coupled directly to the spin of the
atoms through $ \zeta $. From its definition it can be seen that $ \zeta $
is of order $\bit $ so that,
substituting  $ H'_{\rm new}$ in (\ref{eq:hi}), gives the effective
Hamiltonian up to and including terms of order $ \bit \zeta^2 \sim \bit^3 $.
The effects of this type of interaction are considered in the next
paragraph.

\paragraph{Noisy spin interaction}

We now consider 
\beq
H' = \zeta \sigma_+ + \zeta^* \sigma_-,
\eeq
which to lowest order (using (\ref{eq:hi})) gives
\beq
\hef = H_0 - \im \slow{\vartheta^* \dot \vartheta} \sigma_3 ; \qquad
 \zeta = i \dot \vartheta,
\eeq
corresponding to the non-trivial replacement
\beq
 \epsilon \to \epsilon_{\rm eff} = \epsilon - 
\im \slow{\vartheta^* \dot \vartheta} .
\label{eq:eps.eff}
\eeq

This replacement also describes the leading (average) effect of
(\ref{eq:firstnoise}) provided we identify $ \xi = \ddot\vartheta
[ 1 + O ( \bit)] $. If $ \xi \sim \bit^ 0 $ then the modification is
indeed of order $ \bit^3 $. In order to go back to the original problem
we act with $S$ on the states.

The effects of the external sources is, to lowest order, summarized by
the simple shift (\ref{eq:eps.eff}) which
corresponds to a change in the energy gap between the
two spin states of the ``atom'' of this model. In particular, for noise of 
sufficiently large amplitude, we can have $ \epsilon_{\rm eff} = \omega_0
/2 $ in which case the Rabi frequency vanishes, $ E_{n+} = E_{n-} $ and
photon number is conserved.

The shift in $ \epsilon $ can also lead to resonant behavior
between states of different $n$. For
example the energies for the states $ | 0-\rangle $ and $ | 1 -
\rangle $ are equal provided
\beq
\epsilon_{\rm eff} - {1\over2}\omega_0 = { \Omega_0^2 \over 8 \omega_0 } 
\pm {1\over2}\omega_0 
\label{eq:jcres}
\eeq
which has a solution only for $ 0 < \omega_0/\Omega_0 \le 1/2 $. 

These results are reliable provided $ \bit \ll 1 $ which corresponds to
$ | \vartheta | \ll1 $ and $ | \omega_0| ,~ | \Omega_0 | ,~ | \epsilon|
\ll \Omega $. As in the two level system resonances occur when
$|H_0|/\Omega \sim \eta^2 $.

\subsection{Quantum system in an inhomogeneous potential}
\label{sect:exa.al}

The ideas presented in the previous sections can be translated
to the case of a particle whose Hamiltonian is of the form
\beq 
H = - \inv{2m} \nabla^2 + V_0 + V_1,
\eeq
where $V_1$ varies rapidly with position.
For this case we consider the time-independent Schr\"odinger
equation $ H \Psi = E\Psi$ and look for solutions $ \Psi = \psi + \chi$
where $ \chi $ is of small amplitude but exhibits rapid variation in position
while $ \psi $ is slowly varying and of large amplitude.
Substituting this Ansatz we find
\bea
\left(  - \inv{2m} \nabla^2 + V_0 \right) \psi &\simeq& 
E \psi - \slow{V_1 \chi} ,\cr
 - \inv{2m} \nabla^2 \chi &\simeq& -V_1 \psi
\eea
which can be solved to lowest order giving
\bea
\hef\psi &=& E \psi ,\cr
\hef &=&  - \inv{2m} \nabla^2 + V_0 + \slow{ V_1 { 2m \over \nabla^2} V_1 }.
\eea
In this case, for any quantity $ A $, $ \slow{A}$ denotes
the part of $A$ (if any) which varies slowly with position.

In one dimension
the same result can be obtained by converting the time-independent 
Schr\"odinger's equation
to a first order equation for the vector $ ( \Psi, -i d \Psi/d x ) $
and substituting in (\ref{eq:theh}) or (\ref{eq:hscnd}), and using $x$ as
the evolution parameter.

The additional term in $\hef $ is {\em negative} definite and will tend
to bind the particle. In particular, taking $ V_0 = 0 $ and assuming $
V_1$ vanishes at infinity the effective Hamiltonian $ \hef $ will always
exhibit a bound state (of zero angular momentum)
in $\le 2$ dimensions, that is, in $ \le 2 $ dimensions a
rapidly varying potential of zero average will always exhibit localized
states. The same is true in higher dimensions provided the amplitude of
$V_1 $ is large enough

\section{Probabilistic considerations}
\label{sect:fp}

In this section we provide an alternative view of the problem
using the Fokker-Plank equation. For simplicity we consider
the case of a two-level system with Hamiltonian
\bea
H &=& H_0 + H' ,\cr
H_0 &=& \sum_a f^a \sigma_a ,\cr
H' &=& \sum_a \pr^a \sigma_a,
\label{eq:f.p.h}
\eea
where $\{\pr^a\}$ are stochastic variables whose probability
function will be described below.

We will study the Fokker-Plank for the polarization vector
$ \psi^\dagger \sigbf \psi $ whose
probability density is given by
\beq
\pcal(\rr,t) = \avg{\delta\up3\left( \psi^\dagger(t) \sigbf \psi(t) - \rr
\right)}_\pr,
\eeq
where the symbol $\avg{\cdots}_\pr$ denotes the average over
the stochastic variables $\pr^a$. In terms of a functional
integral we will use
\bea
\avg{A}_\pr = \int \prod_a [d\pr^a] \; A \; \exp\left\{
- \half \int\!\int dt \; dt' \sum_{a b}\pr^a(t) \kcal_{ab}(t,t') \pr^b(t')
\right\},
\label{eq:ave}
\eea
with $\kcal $ symmetric ($  \kcal_{ab}(t,t') =  \kcal_{ba}(t',t)$)
and positive definite. We denote by
$\sigma $ the inverse kernel $ \kcal^{-1}$,
\beq
\int ds \sum_c \kcal_{ac}(t,s) \sigma^{cb}(s,t') = \delta^b_a
\delta(t-t').
\label{eq:k.sigma}
\eeq
It is easy to see that $ \sigma^{ab}(t,t') = \avg{\pr^a(t) \pr^b(t')} $.

We will now restrict further considerations to cases where $\pr^a(t)$
is correlated with  $\pr^b(t')$ only for $ t $ close to $t'$, that is for
cases where $ \sigma_{ab}(t,t')$ vanishes except when $ t \sim t' $.
In this case we define
\beq
\bar\sigma_{ab}(t) = \int_{-\infty}^t dt' \; \sigma_{ab}(t,t'),
\label{eq:sigma.bar}
\eeq
and, following the standard derivation of the Fokker-Plank equation~\cite{fp,difuse}, we
obtain,
\beq
i \dot \pcal = \left( 2 \sum_a L_a f^a - 4 i \sum_{ab} L_a L_b
\bar\sigma^{ab} \right) \pcal ,
\label{eq:FPi}
\eeq
where the $f^a$ determine $H_0$ in (\ref{eq:f.p.h}) and $L_a,~a=1,2,3$
denote the usual angular momentum operators in 3 dimensions. There are
corrections to this equation but these can be ignored provided $
\sigma(t,t') $ is sufficiently localized around $ t= t'$.

We will now restrict ourselves to situations where
$ \bar \sigma $ takes the form
\beq
\bar\sigma^{ab} = \half D \delta^{ab} - \half  \sum_c \epsilon^{abc} a_c.
\label{eq:Da}
\eeq
The first term is commonly used in treating this type of problems,
the second term implies a correlation between $\pr^a$ and $\pr^b$ with $ a
\not= b$ and is usually assumed to vanish; we will find, however that it
is precisely this term that is responsible for the resonances described
previously. 

Substituting (\ref{eq:Da}) in (\ref{eq:FPi}) yields
\beq
i \dot \pcal = 2 \left[ ( \ff-\aa)\cdot \LL - i D L^2 \right] \pcal.
\label{eq:FPii}
\eeq 
It is important
to note that this choice still corresponds to a positive definite
kernel $ \kcal $ so that (\ref{eq:ave}) is well defined. 

In order to relate these expressions to the ones obtained previously we 
first write $ \bar \sigma $ in terms of the two-point correlator,
\beq
\bar\sigma^{ab}(t,t') = \int_{-\infty}^t dt' \; \avg{\pr^a(t) \pr^b(t')}_\pr.
\label{eq:Db}
\eeq
We then expand $\pr^a$ in a Fourier series,
\beq
\pr^a(t) = \sum_\omega \pr^a_\omega e^{-\i \omega t}
\label{eq:f.exp.e}
\eeq
(not necessarily restricted to $ | \omega | > \Omega $) and assume
that $ \avg{ \pr^a_\omega \pr^b_{\omega'} }_\pr \simeq 0 $ for $ \omega +
\omega' \not=0 $ (which is equivalent to assuming that the correlator
$\sigma^{ab}(t,t')$ is non-zero for $ t \sim t' $ only). In this case
\bea
D &=& \lim_{\delta\to 0 } {2\over3} \sum_\omega { \delta \over \delta^2 +
\omega^2} \avg{ | \ppr_\omega|^2}_\pr ,\cr
\aa &=&  \lim_{\delta\to 0 } i \sum_\omega 
{\omega  \over \delta^2 +
\omega^2} \avg{\ppr_\omega \times \ppr_{-\omega}})_\pr.
\label{eq:expr.for.D.and.a}
\eea
In obtaining these expressions we included a convergence factor
$e^{\delta t },~\delta \to 0  $ in (\ref{eq:f.exp.e}). 
Note that $D$ will vanish unless
the $ \pr^a_\omega $ are continuously distributed in an interval
containing $ \omega =0 $.

Comparing this result to (\ref{eq:twolevelcase})
we find that the term $\ff-\aa$ in (\ref{eq:FPii})
corresponds to the leading term in $\hef $. In addition, however, the
Fokker-Plank equation contains the non-Hermitian diffusion term $ - 2 i D L^2
\pcal $ which forces
$ \pcal $ to decrease exponentially in time for all but the zero-angular
momentum modes~\cite{difuse}. Qualitatively this implies that for large times the
polarization vector will end up uniformly distributed and the effects of
the $ \ff-\aa $ term will be completely washed-out. For intermediate
times, however, the presence of the $\ff-\aa$ term can lead to
interesting effects.

The general solutions to (\ref{eq:FPii}) can be obtained in terms of 
spherical harmonics~\footnote{It is easy to see that $\pcal $
is, in fact, independent of $|\rr|$.} $ Y_m^l$,
\beq 
\pcal = \sum_{l m} \xi_m^l(t) e^{ - 2 D l(l+1) t - i v_0 m t}
 Y_m^l( \hat \rr),
\label{eq:pexp}
\eeq
where the exponential is introduced for later convenience.
Since $ \pcal $ is real the coefficients $ \xi_m^l $ obey
\beq 
\xi_m^l{}^* = (-1)^m \xi_{-m}^l.
\eeq

The above expansion can be substituted into (\ref{eq:FPii}) leading,
for each $l$, to a set of $ 2l+1$ coupled ordinary differential 
equations in $t$ that can
in principle be solved for any choice of \ff, \aa\ and $D$. 

To
illustrate this procedure we will consider a special case
which is similar to the one often studied when considering 
stochastic resonances. We take
\beq
\ff-\aa = \half v_0 \hat\zz + \half u \cos(\omega_0 t ) \hat\xx,
\label{eq:faexample}
\eeq
corresponding, for example, to $f^3=\pr^3=0$ (so that $a^1=a^2=0$), $f^2 =0$, 
$ f^1 = u \cos(\omega_0 t) $ and $a^3 =- v_0 $. The
equations for the coefficients $ \xi^l_m $ 
in (\ref{eq:pexp}) then become
\beq
\dot \xi^l_m + {i\over2} u \cos(\omega_0 t )  
\left[ 
\sqrt{ l ( l + 1 ) - m ( m - 1 )} e^{  i v_0 t} \xi_{m-1}^l
+ \sqrt{ l ( l + 1 ) - m ( m + 1 )} e^{- i v_0 t} \xi_{m+1}^l
\right] = 0,
\label{eq:xi.dif.eq}
\eeq
For $ l =0 $ the solution is simply $ \xi_0^0 = $constant which is
determines the normalization of $ \pcal $; the equations for $ \l
\not=0 $ can be solved numerically using standard techniques. The case
$ \l=1 $ is of special interest since the coefficients $ \xi_m^{l=1}$
determine the average polarization of the system as a
function of time:
\bea
\avg{\psi^\dagger(t) \sigbf \psi(t) } &=& \int d^2\hat\rr 
\left( \pcal \hat \rr \right) \left/ \int d^2 \hat\rr \pcal \right. \cr 
&=& {\sqrt{2/3}\over \xi_0^0} e^{- 4 D t}
\left( 
 \re\left[\xi_{-1}^1(t) e^{ i v_0 t} \right] , 
 \im\left[\xi_{-1}^1(t) e^{ i v_0 t} \right] , 
\inv{\sqrt{2}} \xi_0^1(t) \right).
\label{eq:av.pol}
\eea

\begin{figure}
\centerline{\vbox to 2.9in{\epsfxsize=11in\epsfbox[-300 -200 962 592]{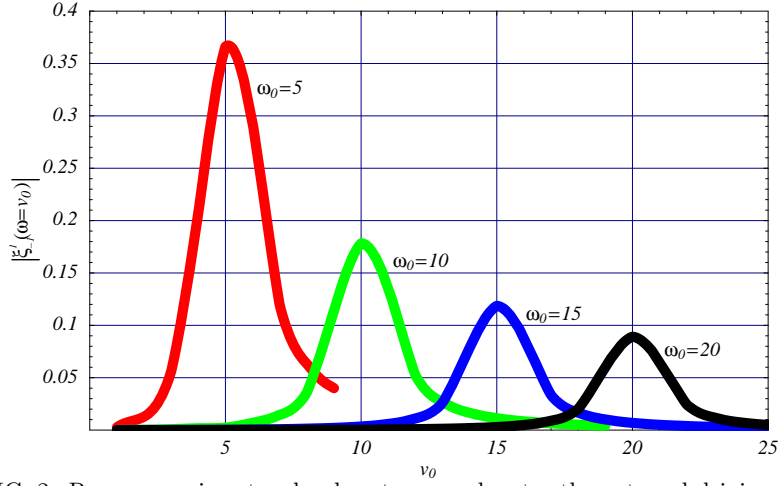}}}
\caption{Resonances in a two-level system; $ \omega_0 $ denotes the external driving
frequency.}
\label{fig:fii}
\end{figure}

\begin{figure}
\centerline{\vbox to 2.9in{\epsfxsize=5 in\epsfbox[0 -200 612 592]{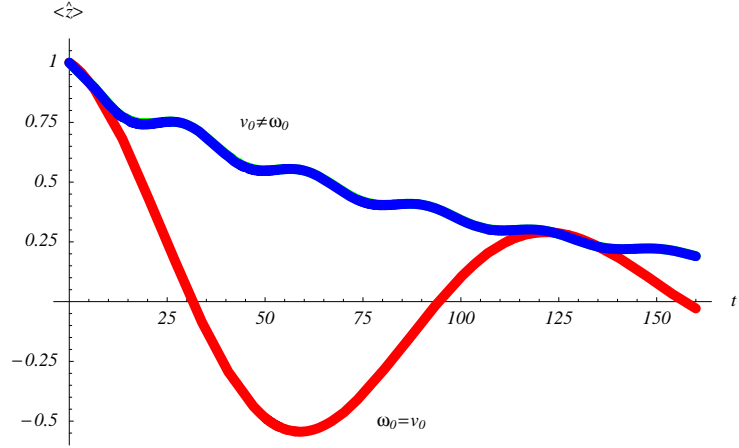}}}
\caption{Resonant behavior of the $z$ component
polarization vector (\ref{eq:av.pol}), $ \left\langle \psi^\dagger
\sigma^3 \psi \right\rangle = \left\langle \hat z \right\rangle $, for
a two level system. The parameters chosen were $ u=0.1$, $\xi_0^0=1/\sqrt{3}$,
$D=0.0025 $, $ \omega_0 =5 $ and, for
the case $ v_0\not=\omega_0 $, $ v_0=\omega_0 \pm 0.2$ }
\label{fig:fiii}
\end{figure}

We will be interested in the possibility of the system resonating at
the driving frequency $ \omega_0 $. This can be investigated by first solving the
above equations and then Fourier transforming the result. We define,
\beq
\tilde \xi_m^l (\omega) = \int dt e^{i \omega t} \xi_m^l (t),
\eeq
and we study the behavior of $ \left| \xi_1^1 (\omega=v_0) \right| $ as
a function of $v_0 $ for various values of $ \omega_0 $ (note that
$\tilde \xi $ also depends explicitly on $v_0 $ since this parameter appears in
the differential equation (\ref{eq:xi.dif.eq})).
The result is presented in Fig. \ref{fig:fii} which clearly shows an
enhancement in the Fourier coefficients of frequency $v_0$ when
$ v_0 = \omega_0 $, the shape of the curves are characteristic of
resonant behavior. These resonances are also illustrated by the behavior of 
$ \avg{\psi^\dagger \sigma^3 \psi } $ for
various values of $v_0, ~ \omega_0 $. 
An example is plotted in fig. \ref{fig:fiii}.

The same system can be studied using the time averaging procedure of
sections \ref{sect:formalism}-\ref{sect:unit.transf}
provided we assume that the corresponding restrictions on the parameters
are satisfied. Assuming this is the case, the effective Hamiltonian corresponding
to the choice (\ref{eq:faexample}) is readily seen to be, to lowest
order in $\bit $,
\beq
\hef = \pmatrix{v_0/2 & \cos(\omega_0 t) \cr \cos(\omega_0 t) & -v_0/2 }.
\label{eq:hef.driven}
\eeq
The corresponding Schr\"odinger equation for the slow modes $ \hef \psi
= i \dot \psi $ can be solved numerically using standard techniques and
the solutions are seen to exhibit resonances  whenever $ v_0 $ is an
integral multiple of $ \omega_0 $. 

There are, of course, differences between the solutions to the
Schr\"odinger equation associated with (\ref{eq:hef.driven}) and the
solutions derived from (\ref{eq:xi.dif.eq}).
The quantities $ \pr^a $ used in obtaining (\ref{eq:xi.dif.eq}) are
assumed to be stochastic variables whose distribution is determined by
(\ref{eq:ave}, \ref{eq:k.sigma}, \ref{eq:sigma.bar}, \ref{eq:Da}). In
contrast, when deriving (\ref{eq:hef.driven}) we assumed the $ \pd^a_\omega $ 
are non-zero only for  $ | \omega| > \Omega $, and we also required
$ \Omega  \gg | \omega_0|, ~|v_0|, ~ |\dot \pd_\omega/\pd_\omega| $.

The values for \aa\ obtained in cases become
identical if we assume that the average over the stochastic variables
in the first case give the same result as the average over time
intervals much larger than $ 1/\Omega$ in the second case. 

The diffusion term $D$ will vanish, as mentioned above, unless the $
\pr_\omega $ are distributed continuously around $ \omega =0 $. This 
term corresponds to the
non-Hermitian contribution (\ref{eq:hscnd.nh}) which, for this case is
simply proportional to the unit matrix. Taking $f^{2,3}=\pd^3=0 $,
in the example of section \ref{sect:exa.n.level} (for the case $N=2$) we
find
\beq
\hef-\hef^\dagger = - i d, \quad d = {d\over dt} \so \left| {
\pd^1_\omega + i \pd^2_\omega \over \omega} \right|^2,
\eeq
which is of order $ \bit^2 $. Note that $d=0$ if
$\pd^{1,2}_\omega $ are time independent; this corresponds to the
vanishing of $D$ should $\pr_\omega $ vanish when $ \omega $ lies in an interval around $
\omega =0 $.

In concluding this section we note that  it is possible to
generate the required correlation between $\pd^1$ and $\pd^2$ by
mixing and filtering two uncorrelated functions $n^{1,2}$. The
details are presented in the Appendix.

\section{Comparison with the standard stochastic resonances}
\label{sect:compare}

The resonances described above are reminiscent of the
well-studied stochastic resonances~\cite{sr} that are characterized by
an increased sensitivity to small perturbations when
noise of an optimal amplitude is introduced. This feature is also
observed when the condition (\ref{eq:res.cond}) is satisfied, and is
illustrated by the behavior of the system studied in the previous section.

More specifically, stochastic resonances occur when there is a match between
a noise-induced transition rate, $ r_N $ and the one produced by an external
perturbation. If the latter is assumed to be harmonic of frequency $
\omega_0 $, then typically resonances occur when
$ \omega_0 \sim \pi r_N $. This can be understood by considering
a system that initially has 2 degenerate minima, such that 
the harmonic perturbation
will first favor one and then the other (alternating with period $
\pi/\omega_0$). If the
resonance condition on $ r_N $ is realized, then the times at which
one minima is disfavored will coincide with the times at which
noise-induced transitions to the other minima are most probable, and
this enhances the response of the system to the external perturbation.
This behavior is also
observed in the systems studied above, for example, the resonances
in Fig. \ref{fig:fii} occur when $ v_0 = \omega_0 $ where, according to 
(\ref{eq:hef.driven}), $v_0 $ is
proportional to the noise-induced transition rate.

There are, however, some technical differences. To illustrate these we
consider the following one-dimensional system
that exhibits stochastic resonances
\beq
\dot x = V'(x,t) + e(t), \quad V(x,t) = V_0(x) + u \; x \; \cos(\omega_0 t),
\eeq
where $e$ denotes a stochastic variable, $u$ is a small coupling constant and $V_0$
is a potential with two (degenerate) minima.
The noise is assumed to obey
\beq
\avg{e(t) e(t') } = F \delta(t-t').
\label{eq:s.r.noise}
\eeq

Following the same steps~\cite{hu} described above it is possible to obtain the
Fokker-Plank equation for the probability density $ \pcal(y,t) =
\avg{\delta\left( x(t) - y \right)}_e $ and the corresponding average
$ \avg{ x(t) }_e $. The Fourier coefficient of $ \avg{ x(t) }_e $
corresponding to frequency $ \omega_0 $ has an amplitude proportional to
$ (\lambda/F) /\sqrt{\omega_0^2 + \lambda^2}$ 
where $ \lambda $ denotes the noise-induced
hopping rate (the Kramers' rate~\cite{kr}), $ \ln \lambda \propto -1/F $.  For
fixed $ \omega_0 $ this
amplitude also displays an enhancement at a certain value of $ F $~\cite{sr}. 
Comparing these results with those obtained in the previous section we
note that
\begin{itemize}
\item The usual stochastic resonances occur for uncorrelated noise
obeying (\ref{eq:s.r.noise}) while resonance behavior in (\ref{eq:f.p.h})
requires the correlations implied by having $\aa\not=0$ in (\ref{eq:Da}).
\item The resonances described above have the usual shape
(see Fig. \ref{fig:fii}) for
the resonant curve. This is not necessarily the case for the stochastic
resonances usually discussed in the literature.
\item Usual stochastic resonances occur whenever the driving frequency
is about half the Kramers' rate, which depends exponentially on
the noise level $F$. For the case
presented in this paper resonances occur when the driving frequency is
$\sim |\aa|$ as defined in (\ref{eq:expr.for.D.and.a}), and is
proportional to the square of the amplitude of the stochastic variables
$\pr^a$.
\end{itemize}

\section{Conclusions}
\label{sect:concl}

In this work we propose a formalism
which makes possible to study systems under the influence of 
rapidly-varying external fields (which are not necessarily perturbative) whose
typical frequency we denoted by $ \Omega $.
The formalism provides a solution as a power series in $ 1/\Omega $ and
assumes a clear separation of fast (frequencies $ \gesim \Omega $) and slow
(frequencies $ \ll \Omega $) modes. 

We showed that the evolution of the
slow modes is determined by an effective Hamiltonian which is not
necessarily Hermitian; a point noted in other related
calculations~\cite{bur}. The non-Hermitian contributions to the
effective Hamiltonian, however, can be eliminated by performing an
appropriate unitary transformation.

The formalism was applied to various classical and quantum systems. 
In some examples we
found that the external field can produce a resonant
behavior in the system. These resonances are related, but not identical, to
the stochastic resonances studied in the literature~\cite{sr} 
In particular the resonant phenomena studied in this paper occur only
when the interaction with the environment involves several correlated
terms.

In one particular application of the formalism we argued that the presence of a
random time-independent potential will necessarily generate bound states
in systems of dimension $1$ and $2$, and in other dimensions as well
provided the amplitude of the potential is sufficiently large. The
connection of this result with the phenomenon of Anderson localization~\cite{al}
are tantalizing and will be considered in a future publication.

\acknowledgements
We would like to thank 
W. Beyermann,  R. deCoss and T.J. Weiler
for illuminating comments and insights.
This research was supported in part by US DOE contract 
number DE-FG03-94ER40837(UCR) and by Conacyt (M\'exico).

\section*{Appendix}

In this appendix we describe a simple construction which generates
stochastic variables $\pd^a$ satisfying (\ref{eq:Da},\ref{eq:Db}), in 
terms of a set of uncorrelated variables $ n^i$, specifically,
we assume
\beq
\avg{n_i(t) n_j(t')} = \half D_i \delta_{ij} \delta(t-t'),
\eeq
and search for new variables $\pd^a$ satisfying
\beq
\avg{\pd^a(t) \pd^b(t')} = \half F \ecal(t-t') \delta_{ab} - \half
\epsilon_{abc} a^c \ocal(t-t'),
\eeq
where $ \ecal $ is an even function of its argument while $ \ocal $
is odd and satisfy
\beq
\int_{-\infty}^0 \ecal(s) ds = \int_{-\infty}^0 \ocal(s) ds = 1.
\eeq
In terms of the Fourier transformed quantities,
\beq
\tilde \pd^a(\omega) = \int_{-\infty}^\infty dt \; e^{ - i \omega t } \pd^a(t), \quad
\tilde n_i(\omega) = \int_{-\infty}^\infty dt \; e^{ - i \omega t } n_i(t),
\eeq
we require
\bea
\avg{\tilde \pd^a(\omega) \tilde \pd^b(\omega') }
&=& \pi \delta(\omega+\omega')\left[ F \delta_{ab} \tilde\ecal(\omega) 
- \epsilon_{a b c} a^c \tilde \ocal(\omega) \right] ,\cr  
\avg{\tilde n_i(\omega) \tilde n_j(\omega') } 
&=& \pi \delta(\omega+\omega')D_i \delta_{ij},
\eea
where
\bea
&& \tilde\ecal(\omega)^*=\tilde\ecal(-\omega)=+\tilde\ecal(\omega),\cr
&& \tilde\ocal(\omega)^*=\tilde\ocal(-\omega)=-\tilde\ocal(\omega).
\eea

We look for a linear relation between $ \tilde \pd^a $ and $\tilde  n_i $, namely
\beq
\tilde \pd^a(\omega) = \sum_i K_{ai}(\omega) n_i(\omega).
\eeq
Writing 
\beq
K_{ia}=\inv{\sqrt{ D_a F \tilde \ecal}}
\left(Q_+ + Q_-\right)_{ia},
\eeq
with $ Q_\pm(\pm\omega) = \pm Q_\pm(\omega ) $
and assuming $ Q_\pm^T = \pm Q_\pm $ (where $T$ indicates the transpose)
we find
\beq
Q_+^2 + Q_-^2 = 1, \qquad \{Q_+,Q_-\}_{ij} = \epsilon_{ikj} \nu^k,
\eeq
where $ \nu^k = a^k \tilde \ocal/(F \tilde \ecal) $. These equations
are solved, for example, by choosing
\beq 
Q_+ = \cosh u \left( 1+ \hat\aa\otimes\hat\aa \right), \qquad (Q_-{})_{ij} = 
\sinh u \epsilon_{ikj}\hat a^k
\eeq
with $ \sinh(2u) = | \nubf|$.

It follows that given a set of uncorrelated variables $ n_i$ it is possible
to generate the desired correlated quantities $ \pd^a$ through a linear
filter defined by the (frequency-dependent) matrix $K$.

\end{document}